\documentclass[11pt,twoside]{article}
\usepackage{asp2010}
\resetcounters

\begin{document}
\title{Relativistic Hydrodynamic Codes for Adiabatic and Isothermal Flows}

\author{Dongsu Ryu$^1$, Indranil Chattopadhyay$^2$, and Hanbyul Jang$^1$}
\affil{$^1$Department of Astronomy and Space Science, Chungnam National
University, Daejeon 305-764, South Korea\\
$^2$ARIES, Manora Peak, Nainital-263129, Uttaranchal, India}

\begin{abstract} 

The equation of state (EOS) is an important issue in numerical simulation
codes for relativistic hydrodynamics.
We describe a code for adiabatic flows, employing an EOS which is simple
and yet approximates very closely the EOS of perfect gas in relativistic
regime.
We also describe a code for isothermal flows, where the EoS is trivially
given.

\end{abstract}

\section{Introduction} 

Highly energetic phenomena, which are relativistic in nature, are common
in astrophysical environments:
accretion disks around black holes
\citep[see, e.g.,][for review]{mill07},
relativistic jets from Galactic sources
\citep[see, e.g.,][for review]{mir99},
extragalactic jets from active galactic nuclei
\citep[see, e.g.,][for review]{zen97},
and gamma-ray bursts \citep[see, e.g.,][for review]{mes02}.
Gas in such relativistic phenomena is characterized by its relativistic
fluid speed ($v \rightarrow c$) and/or relativistic sound speed
($c_s \rightarrow c/\sqrt{3}$).

Numerical codes for relativistic hydrodynamics (RHDs) have been
successfully built, based on schemes that were originally developed for
codes for non-relativistic hydrodynamics.
Most codes employed the equation of state (EoS) of the gas, which was
designed for the gas with a constant ratio of specific heats and
so is essentially valid only for the gas in either non-relativistic
or ultra-relativistic regime \citep[see, e.g.,][for reviews]{mar03,wil03}.
The correct EoS (see Section 3), however, involves the specific enthalpy
expressed in terms of the modified Bessel functions \citep[see][]{syn57}.
While codes employing the correct EoS has been built
\citep[see, e.g.,][]{fk96,samg02}, they normally comes with an extra cost
of computation time.
On other other hand, approximate EoSs that mimic the correct EoS have
been suggested \citep[see, e.g.,][]{math71,serv86,rcc06}, and recently,
codes employing those approximate EoSs have been introduced
\citep[see, e.g.,][]{mpb05,rcc06}.

In this paper, we describe a RHD code for adiabatic flows, which
was presented in \citet{rcc06};
an EoS, which is simple and an algebraic function of temperature,
was employed.
We also describe a new RHD code for isothermal flows;
the meaning of isothermality in relativistic regime is discussed.
The steps necessary to build codes including the transformation from
the conserved quantities to the primitive quantities and the
eigen-structure are presented.
Finally, shock tube tests performed with codes based on the Total
Variation Diminishing (TVD) scheme are presented.

\section{Relativistic Hydrodynamic Equations}

The special RHD equations for an ideal fluid in the laboratory frame
of reference can be written as
\begin{equation}
\frac{\partial D}{\partial t}+\frac{\partial}{\partial x_j}
\left(Dv_j\right) = 0,
\end{equation}
\begin{equation}
\frac{\partial M_i}{\partial t}+\frac{\partial}{\partial x_j}
\left(M_iv_j+p\delta_{ij}\right) = 0,
\end{equation}
\begin{equation}
\frac{\partial E}{\partial t}+\frac{\partial}{\partial x_j}
\left[\left(E+p\right)v_j\right] = 0,
\end{equation}
where $D$, $M_i$, and $E$ are the mass density, momentum density, and
total energy density, respectively \citep[see, e.g.,][]{ll59}.
The conserved quantities in the laboratory frame are expressed as
\begin{equation}
D = \Gamma\rho, \qquad M_i =\Gamma^2{\rho}hv_i, \qquad
E = \Gamma^2{\rho}h-p,
\end{equation}
where $\rho$, $v_i$, $p$, and $h$ are the proper mass density, fluid
three-velocity, isotropic gas pressure and specific enthalpy,
respectively, and the Lorentz factor is given by
\begin{equation}
\Gamma = \frac{1}{\sqrt{1-v^2}}\qquad{\rm with}\qquad
v^2 = v_x^2+v_y^2+v_z^2.
\end{equation}
Here, $h \equiv (e+p)/\rho$, where $e$ is the sum of the proper
internal and rest-mass energy densities.
In the above, the Latin indices (e.g., $i$) represents spatial
coordinates and the conventional Einstein summation is used.
The speed of light is set to unity ($c\equiv1$) throughout this paper.

\section{Code for Adiabatic Flows} 

Equations (1) -- (3) with the EoS, $h = h(p,\rho)$, form a hyperbolic
set of conservation equations for adiabatic flows. 
The correct EoS for the single-component perfect gas in relativistic
regime (hereafter RP) can be derived \citep[see][]{syn57}; it is given as
\begin{equation}
h=\frac{K_3(1/\Theta)}{K_2(1/\Theta)},
\end{equation}
where $K_2$ and $K_3$ are the modified Bessel functions of the second
kind of order two and three, respectively.
Here, $\Theta = p/{\rho}$ is a temperature-like variable.
Using the EoS in (6), however, poses a difficulty, because the inverse,
that is, $\Theta$ as a function of $h$, can not be expressed as a
simple form.

\begin{figure}[!ht]
\vskip -2.5cm\hskip 0.3cm
\includegraphics[width=0.95\textwidth]{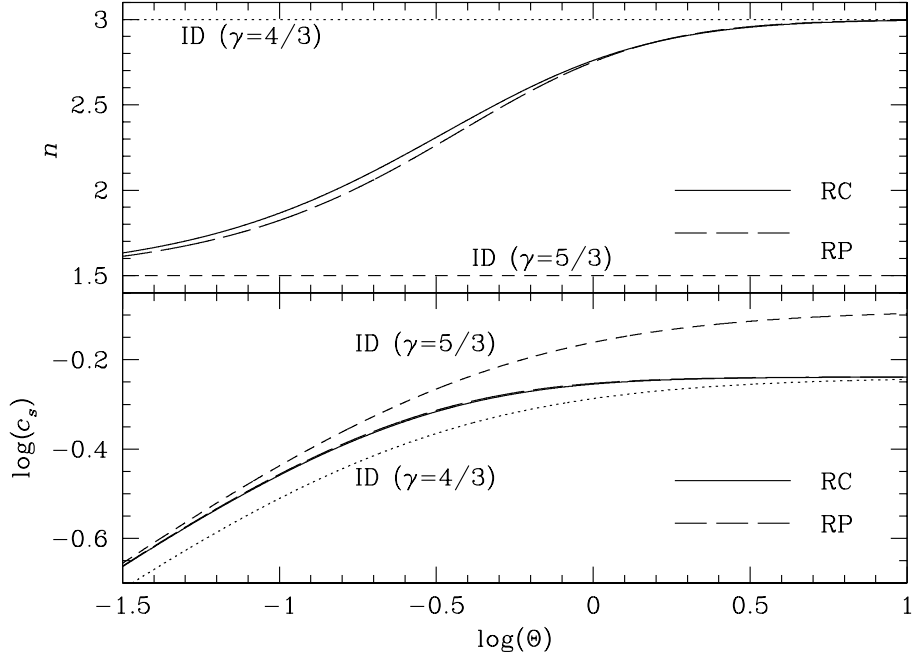}
\vskip -1.5cm
\caption{Polytropic index, $n$, and sound speed, $c_s$, as a function of
temperature, $\Theta = p/{\rho}$, for adiabatic flows employing different EoSs.}
\end{figure}

\citet{rcc06} introduced an approximate EoS
(hereafter RC),
\begin{equation}
h=2\frac{6\Theta^2+4\Theta+1}{3\Theta+2}.
\end{equation}
RC mimics very closely RP.
Figure 1 compares the polytropic index and sound speed
\begin{equation}
n = {\rho}\frac{{\partial}h}{{\partial}p}-1, \qquad
c_s^2 = -\frac{{\rho}}{nh}\frac{{\partial}h}{{\partial}{\rho}},
\end{equation}
for RP and RC as well as for the EoS with a constant ratio of specific
heats, $\gamma$, (hereafter ID, standing for ideal gas).
The specific enthalpy $h$ for RC fits $h$ for RP within the error of
0.8 \%.

Building codes based on upwind schemes requires the eigen-structure
(eigenvalues and eigenvectors) for the relevant hyperbolic set of
conservation equations.
The eigen-structure for RHD equations for adiabatic flows in (1) -- (3)
are given in \citet{rcc06}.
The eigenvalues are
\begin{equation}
a_1 = \frac{\left(1-c_s^2\right)v_x-c_s/\Gamma\cdot\sqrt{Q}}{1-c_s^2v^2},
\end{equation}
\begin{equation}
a_2 = a_3 = a_4 = v_x,
\end{equation}
\begin{equation}
a_5 = \frac{\left(1-c_s^2\right)v_x+c_s/\Gamma\cdot\sqrt{Q}}{1-c_s^2v^2},
\end{equation}
where $Q=1-v^2_x-c^2_s(v^2_y+v^2_z)$, assuming that the flow varies
along the $x$-direction.
For the right and left eigenvectors, refer \citet{rcc06}.

\begin{figure}[!ht]
\vskip -1cm\hskip 0.3cm
\includegraphics[width=0.95\textwidth]{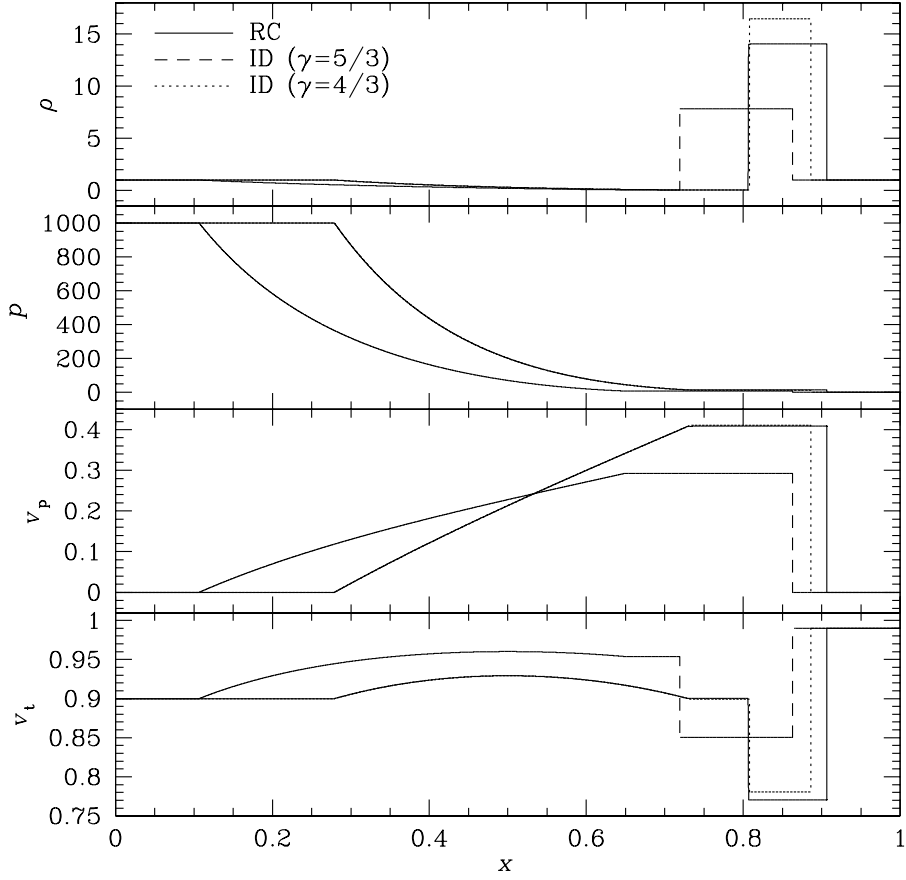}
\vskip -0.2cm
\caption{A relativistic shock tube from the code for adiabatic flows employing
different EoSs.}
\end{figure}

Equations (1) -- (3) evolve the conserved quantities $D$, $M_i$, and $E$,
but the primitive quantities, $\rho$, $v_i$, and $p$, are necessary to
calculate the eigenvalues and eigenvectors.
Combining (4) -- (5) along with (7) results in
\begin{eqnarray}
& M\sqrt{\Gamma^2-1}\left[3E\Gamma(8\Gamma^2-1)+2D(1-4\Gamma^2)\right]
\nonumber \\
& = 3\Gamma^2\left[4(M^2+E^2)\Gamma^2-(M^2+4E^2)\right]
-2D(4E\Gamma-D)(\Gamma^2-1).
\end{eqnarray}
Further simplification reduces this to an equation involving the
$8^{\rm th}$ power of $\Gamma$.
The above can be solved numerically to get $v$.
And then, the rest of the primitive quantities can be solved.

\begin{figure}[!ht]
\vskip -1cm\hskip 0.3cm
\includegraphics[width=0.95\textwidth]{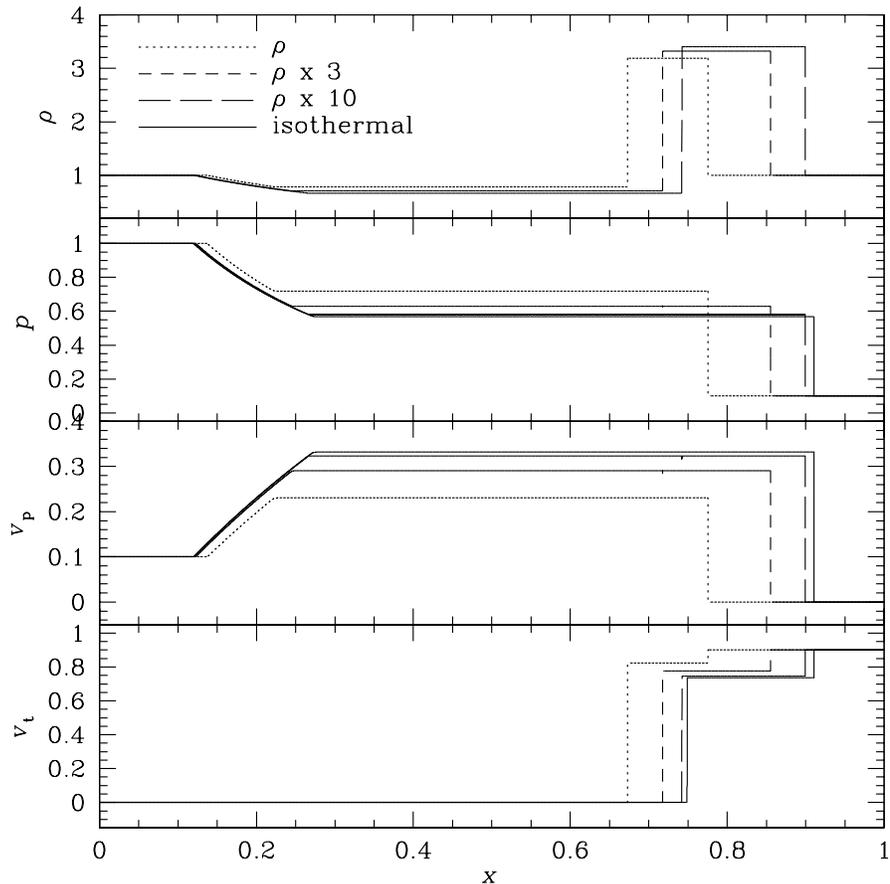}
\vskip -0.2cm
\caption{A relativistic shock tube from the code for isothermal flows
and from the code for adiabatic flows employing RC with different initial
densities, $\rho = 1$, 1/3, 1/10.}
\end{figure}

A RHD code for adiabatic flows was built based on the TVD scheme
\citep{rcc06}.
Figure 2 show a shock tube test comparing the results with RC
and ID: initially $\rho_L = \rho_R=1$, $p_L=10^3$, $p_R=10^{-2}$,
$v_{p,L} = v_{p,R} = 0$, $v_{t,L}=0.9$, $v_{t,R}=0.99$, and
$t_{\rm end} = 0.75$.
Here, the subscripts $L$ and $R$ denote the quantities in the left
and right states of the initial discontinuity at $x=0.5$, and
$t_{\rm end}$ is the time when the solutions are presented.
And $v_p$ and $v_t$ are the velocity components parallel and transverse
to the propagation of structures (i.e., the $x$-direction).
The ID solution with $\gamma = 4/3$ matches well to the RC solution
in the left of the contact discontinuity where the flow has
$\Theta \gg 1$.
But a difference is obvious in the region between contact discontinuity
and shock because $\Theta \sim 1$ there.
The ID solution is clearly different from the RC solution, indicating
the importance of using correct EoSs.

\section{Code for Isothermal Flows} 

As in non-relativistic hydrodynamics \citep[see, e.g.,][]{krjh99},
a code for isothermal flows, where the EoS is given by
$p = c_s^2\ e$ with a constant sound speed $c_s$, can be built in RHDs.
Such EoS arises in several important situations.
Most of all, when the constituent particles are ultra-relativistic or
the fluid is dominated by radiation, that is, $e \gg \rho$, the sound
speed goes $c_s \rightarrow 1/\sqrt{3}$ and the EoS becomes $p = (1/3) e$.
Also the EoS in degenerate matters, such as in white dwarfs and neutron
stars, may be modeled as $p = c_s^2\ e$ \citep[see, e.g.][]{wein72}.

Equations (2) -- (3) with $p = c_s^2\ e$ form a complete, hyperbolic
set of conservation equations for isothermal flows. 
The eigenvalues have the same form as those for adiabatic flows in
(9) -- (11), except two $v_s$ instead of three;
this is expect by considering the nature of weak solutions of initial
value problems in RHDs \citep[see, e.g.][]{st93}.
The left and right eigenvectors are substantially simpler than those
for adiabatic flows.
(They will be published elsewhere due to the page limit of this
proceeding paper.)
The calculation of the primitive quantities from the conserved quantities
is also substantially simpler;
\begin{equation}
v = \frac{-\sqrt{\left(1 + c_s^2\right)^2 E^2 - 4 c_s^2 M^2} +
(1 + c_s^2) E}{2 c_s^2 M},
\qquad
p = \frac{M}{v} - E.
\end{equation}
Each component of velocity can be calculated with $v_i = (M_i/M) v$.
Again a RHD code for isothermal flows was built based on the TVD scheme.
Figure 3 shows a shock tube test comparing the results from the code
for isothermal flows and the code for adiabatic flows with RC:
initially $p_L=1$, $p_R=0.1$, $v_{p,L}=0.1$, $v_{p,R}=0$, $v_{t,L}=0$,
$v_{t,R}=0.9$, and $t_{\rm end} = 0.75$.
For the code for adiabatic flows, $\rho_L = \rho_R=1$, $1/3$, and $1/10$.
The adiabatic solution approaches the isothermal solution, as
$\rho \rightarrow$ small or $\Theta \rightarrow$ large.
Our test indicates that for $\Theta \ga$ a few $\times 10$, the two
solutions for isothermal and adiabatic flows become indistinguishable.

In principle, isothermal flows can be simulated with codes for adiabatic
flows, as well.
However, there are advantages of using codes for isothermal flows:
1) Codes for isothermal flows are faster than those for adiabatic flows,
because the eigenvectors are simpler.
With our codes based on the TVD scheme, the isothermal version is about
1.5 to 2 times faster than the adiabatic version.
2) Codes for isothermal flows should be numerically more robust than
those for adiabatic flows, because one mode, which is the entropy mode,
is less.
Our tests have shown that shocks and discontinuities of $v_t$ are better
resolved with the code for isothermal flows.
In addition, we expect that the numerical dissipation would be smaller
in the code for isothermal flows (not shown here).

\acknowledgements

We thank the referee for comments.
The work was supported in part by National Research Foundation of Korea
through grant KRF-2007-341-C00020.

\end{document}